\documentclass[sigconf,table]{acmart}

\renewcommand\footnotetextcopyrightpermission[1]{} 

\AtBeginDocument{%
  \providecommand\BibTeX{{%
    \normalfont B\kern-0.5em{\scshape i\kern-0.25em b}\kern-0.8em\TeX}}}

\acmSubmissionID{272}

\copyrightyear{2022}
\acmYear{2022}
\setcopyright{rightsretained}
\acmConference[SIGGRAPH '22 Conference Proceedings]{Special Interest Group on Computer Graphics and Interactive Techniques Conference Proceedings}{August 7--11, 2022}{Vancouver, BC, Canada}
\acmBooktitle{Special Interest Group on Computer Graphics and Interactive Techniques Conference Proceedings (SIGGRAPH '22 Conference Proceedings), August 7--11, 2022, Vancouver, BC, Canada}
\acmDOI{10.1145/3528233.3530706}
\acmISBN{978-1-4503-9337-9/22/08}

\setcitestyle{square,nosort}
\makeatletter
\renewcommand*{\NAT@spacechar}{~}
\makeatother

\usepackage{soul}
\usepackage{microtype}
\usepackage{amsmath}
\usepackage{amsfonts}
\usepackage{booktabs}
\usepackage{hyperref}
\usepackage{multirow}
\usepackage{graphicx}
\usepackage{dsfont}
\usepackage{pifont}
\usepackage{xcolor}
\usepackage{wrapfig}
\usepackage{nicefrac}
\usepackage{ctable}
\usepackage{natbib}
\usepackage{xspace}
\usepackage{siunitx}

\newcommand{\mywfigure}[3]{%
\begin{wrapfigure}{r}{#2\columnwidth}%
  \begin{center}%
    \includegraphics[width=#2\columnwidth]{\figurePath#1}%
    \\[-1.5ex]
    \caption{#3}%
    \label{fig:#1}%
  \end{center}%
\end{wrapfigure}%
}

\soulregister\cref7
\soulregister\cite7
\soulregister\cite7
\soulregister\shortcite7
\soulregister\eg0
\soulregister\ie0
\soulregister\etal0

\newcommand{\mysection}[2]{\section{#1}\label{sec:#2}}
\newcommand{\mysubsection}[2]{\subsection{#1}\label{sec:#2}}

\newcommand{\refSec}[1]{Sec.~\ref{sec:#1}}

\newcommand{\refFig}[1]{Fig.~\ref{fig:#1}}
\newcommand{\refEq}[1]{Eq.~\ref{eq:#1}}
\newcommand{\refEqs}[2]{Eqs.~\ref{eq:#1} and \ref{eq:#2}}
\newcommand{\refTab}[1]{Tab.~\ref{tab:#1}}

\newcommand{\ie}{i.e.,\ }
\newcommand{\eg}{e.g.,\ }

\newcommand{\mymath}[2]{\newcommand{#1}{\TextOrMath{$#2$\xspace}{#2}}}

\newcommand{\unsure}[1]{{\sethlcolor{yellow}\hl{#1}}}

\definecolor{colorA}{HTML}{4285f4} \definecolor{colorB}{HTML}{ea4335} \definecolor{colorC}{HTML}{fbbc04} \definecolor{colorD}{HTML}{34a853} \definecolor{colorE}{HTML}{ff6d01}
\definecolor{colorF}{HTML}{46bdc6}
\definecolor{colorG}{HTML}{000000}
\definecolor{colorH}{HTML}{777777}

\usepackage[nolist]{acronym}

\begin{acronym}
\acro{2AFC}{two-alternative forced choice}
\acro{EA}{emission-absorption}
\acro{CT}{computed tomography} 
\acro{IoR}{index of refraction}
\acro{NeRF}{neural radiance field}
\acro{NN}{neural network}
\acro{CNN}{convolutional neural network}
\acro{MC}{Monte Carlo}
\acro{MPI}{multi-plane images}
\acro{MLP}{multi-layered perceptron}
\acro{NVS}{novel-view synthesis}
\acro{ODE}{ordinary differential equation}
\acro{RTE}{radiative transfer equation}
\acro{SDF}{signed distance field}
\acro{PSNR}{peak signal-to-noise ratio}
\acro{SSIM}{structural similarity}
\acro{LPIPS}{learned perceptual image patch similarity}
\end{acronym}

\def\figurePath{images/}

\makeatletter
\def\myparagraph{\@startsection{paragraph}{4}{\parindent}%
  {2pt}
  {-\parindent}
  {\ACM@NRadjust{\@parfont\@adddotafter}}}
\makeatother

\def\myfigure#1#2{\begin{figure}[ht]\centering\includegraphics*[width = \linewidth]{\figurePath#1}\\[-1.5ex]\caption{#2}\label{fig:#1}\end{figure}}
\def\mycfigure#1#2{\begin{figure*}[t]\centering\includegraphics*[clip, width = \linewidth]{\figurePath#1}\\[-1.5ex]\caption{#2}\label{fig:#1}\end{figure*}}

\soulregister\ref7
\soulregister\cite7
\soulregister\citeetal7
\soulregister\refSec7
\soulregister\refFig7
\soulregister\refTbl7
\soulregister\refEq7
\soulregister\cite7
\soulregister\ref7
\soulregister\pageref7
\soulregister\shortcite7
\soulregister\eg7
\soulregister\ie7
\soulregister\etal7
\soulregister\unsure7

\DeclareGraphicsExtensions{.png,.jpg,.pdf,.ai,.psd}
\begin{document}

\title{Eikonal Fields for Refractive Novel-View Synthesis}

\author{Mojtaba Bemana}
\orcid{0000-0003-0202-6266}
\affiliation{%
	\institution{MPI Informatik}
	\country{Germany}
}
\email{mbemana@mpi-inf.mpg.de}

\author{Karol Myszkowski}
\orcid{0000-0002-8505-4141}
\affiliation{%
	\institution{MPI Informatik}
 	\country{Germany}
}
\email{karol@mpi-inf.mpg.de}

\author{Jeppe Revall Frisvad}
\orcid{0000-0002-0603-3669}
\affiliation{%
	\institution{Technical University of Denmark}
	\country{Denmark}
}
\email{jerf@dtu.dk}

\author{Hans-Peter Seidel}
\affiliation{%
	\institution{MPI Informatik}
 	\country{Germany}
}
\email{hpseidel@mpi-inf.mpg.de}

\author{Tobias Ritschel}
\affiliation{%
	\institution{University College London}
 	\country{UK}
}
\email{t.ritschel@ucl.ac.uk}

\renewcommand{\shortauthors}{Mojtaba Bemana, et al.}

\mymath{\ior}{n}
\mymath{\location}{\mathbf p}
\mymath{\motion}{\mathbf v}
\mymath{\direction}{\omega}
\mymath{\radiance}L
\mymath{\absorption}{\sigma}
\mymath{\emission}q

\mymath{\ud}{\mathrm d}
\mymath{\optpath}{\mathcal S}
\mymath{\origin}{\mathbf o}
\mymath{\raypos}{\mathbf p}
\mymath{\momentum}{\mathbf v}

\newcommand{\dd}[2]{\frac{\mathrm d #1}{\mathrm d #2}}
\newcommand{\dds}[1]{\dd{#1}{s}}

\newcommand{\revised}[1]{#1}

\begin{abstract}
We tackle the problem of generating novel-view images from collections of 2D images showing refractive and reflective objects.
Current solutions assume opaque or transparent light transport along straight paths following the emission-absorption model.
Instead, we optimize for a field of 3D-varying \ac{IoR} and trace light through it that bends toward the spatial gradients of said \ac{IoR} according to the laws of \emph{eikonal} light transport.
\end{abstract}



\begin{teaserfigure}
   \ifthenelse{\equal{\revised{}}{}}{}{\vspace{1.4cm}}
   \includegraphics[width=\textwidth]{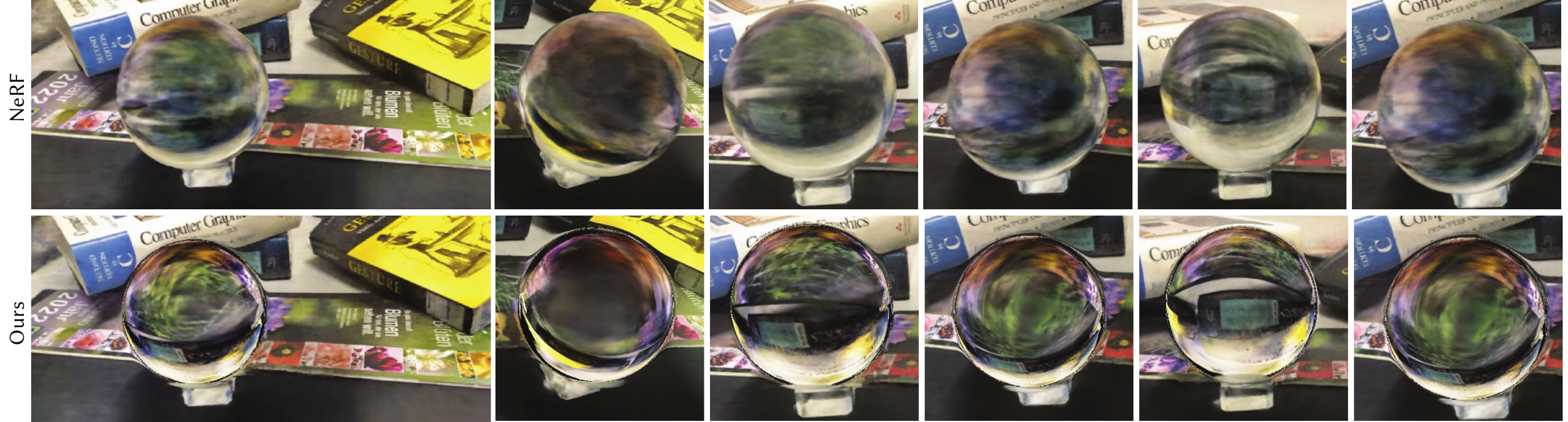}
   \\[-4.5ex]
   \caption{
    Novel-view synthesis using \acl{NeRF} (top) and our eikonal approach (bottom) for a \revised {real refractive} scene.
   }
   \label{fig:Teaser}
\end{teaserfigure}

\maketitle

\acresetall

\thispagestyle{empty}
\pagestyle{plain}

\mysection{Introduction}{Introduction}
Given images \revised{with different views} of a refractive object, it is a challenging task to synthesize a novel view.
The issue is that the refractive object takes its appearance from the surroundings by bending and internally reflecting the rays of light that travel through the object.
If we fully digitize the object and its surroundings, we can synthesize novel views~\cite{trifonov2006tomographic,hullin2008fluorescent,ihrke2010transparent,stets2017scene}, but this approach requires a lot more information than a simple set of images.
We would for instance need a dedicated hardware setup to digitize a transparent object~\cite{ihrke2010transparent,stets2017scene,lyu2020differentiable}.
Deep learning offers an alternative approach where we can instead render transparent objects and train a \acl{NN} to estimate the shape of such objects in more arbitrary surroundings~\cite{stets2019single,sajjan2020clear,li2020through}.
A deep learning technique based on a synthetic dataset, however, often returns a faulty estimate when presented with an image significantly different from those in the training data \cite{lyu2020differentiable}.

We would like to avoid the difficulties in representing a wide enough range of transparent object appearances in one synthetic dataset. One way to do this is to learn the radiance field of a given object based on a set of images capturing its appearance as observed from different directions~\cite{lombardi2019neural,mildenhall2020nerf}. This is useful for locating and estimating the distance to transparent objects~\cite{ichnowski2021dex}. However, since neural radiance fields do not consider refraction, we can not use it out of the box for refractive \ac{NVS}.

To enable this, we devise a method that optimizes for the field of 3D spatially-varying \ac{IoR} given a set of 2D images picturing a refractive object.
Existing solutions to learn 3D fields capturing scene geometry are based on opaque or transparent light transport along straight paths.
In the presence of transparent objects, however, light bends, \ie it changes its direction.
The precise way in which light paths are curved depends on a certain \emph{eikonal equation} operating on spatial gradients of the \ac{IoR} field, which we show can be solved -- and differentiated over in learning -- in practice with the appropriate formulation.
The resulting method allows for novel-view synthesis (Fig.~\ref{fig:Teaser}) in 3D scenes with complex objects involving strong refractive and internal reflection effects. \revised{Our code and training data are publicly available at \url{https://eikonalfield.mpi-inf.mpg.de/}}.

\mysection{Related work}{RelatedWork}
As our focus is novel-view synthesis for refractive transparent objects, we discuss this problem with an emphasis on recent neural rendering solutions that can handle specular effects (\refSec{RWNovelViewSynthesis}).
We overview also image-based modeling of transparent objects (\refSec{RWTransparentSurface}), which is a more general setup than we require in this work, but still some similarities can be found.
Finally, we discuss physics-based eikonal rendering that inspired our work (\refSec{RWEikonalRendering}).

\mysubsection{Novel-view synthesis}{RWNovelViewSynthesis}
Recent learning-based approaches allow synthesizing images under new views without accurately reconstructing the physical parameters \cite{tewari2020state,tewari2021advances}.

\myparagraph{\Acp{NeRF}}
\citet{mildenhall2020nerf} introduce a volumetric opacity representation that encodes both geometry and appearance using a \ac{MLP} and is trained on a large set of multiple-view RGB images and proves to be extremely successful in novel-view synthesis.
More specifically, view-dependent RGB color and view-independent density are learned as sharp functions in space and smooth functions in angle.
In the case of near-mirror or near-glass reflection/refraction, appearance cannot be described as a smooth function of angle anymore  \cite{guo2021nerfren}.
Consequently, reflected or refracted patterns appear notoriously ghosty and blurry \cite{ichnowski2021dex}.
A number of solutions exist \cite{zhang2021NeRFactor,boss2021nerd} that disentangle normal vectors and spatially-varying reflectance by manipulating the \ac{NeRF} density representation, but highly specular surfaces with clearly visible reflected environment cannot be reproduced.
Our approach does not fully rely on the NeRF geometry and uses the diffuse scene only as a backdrop.

Inspired by traditional image-based rendering \cite{sinha2012image,xu2021scalable} for scenes with planar reflections, \citet{guo2021nerfren} introduce NeRFReN, where an additional \ac{NeRF} structure is proposed that renders a reflected image and composes it additively with the traditional \ac{NeRF} rendering.
MirrorNeRF \cite{wang2021mirrornerf} employs a catadioptric imaging system based on an array of hemispherical mirrors enabling a single-shot portrait reconstruction and rendering.
The position of a sample point is warped while its view direction is unchanged. In our approach, the view directions bend.

\myparagraph{Other volumetric representations}
\citet{lombardi2019neural} and others \cite{yu2021plenoxels,yu2021plenoctrees} propose a voxel grid with interpolation optimized using a \acs{CNN} (or a gradient method directly) that encodes both geometry (possibly dynamic) and appearance.
\citet{lombardi2019neural} perform learned warping to reduce memory requirements and to improve the resolvable details.
We instead warp space according to physical laws that regularize the problem.
Even sharper mirror reflection and transparency effects can be obtained using an extension of the \ac{MPI} representation, where for every pixel in a stack of semi-transparent planes, directional information using learned basis functions is stored
\cite{wizadwongsa2021NeX}.
Similarly, as for other \ac{MPI}-based and neural light-field methods \revised{\cite{bemana2020xfields,attal2021learning}}, only narrow baselines are supported.

A \acl{SDF}, possibly encoded into an \ac{MLP}, \revised{can} represent surface geometry and \revised{help recovering} non-spatially-varying reflectance using spherical Gaussians that in turn enable good quality reflections \cite{zhang2020physg}.
In an alternative point-based representation \cite{kolos2020transpr}, where each point is associated with a learnable photometric, geometric, and transparency descriptor, relatively sharp depiction of semi-transparent objects is achieved, when the non-distorted background is also known.
However, specular effects are explicitly excluded from the training data. 

In all those solutions, an important limiting factor is a straight-path assumption in the rendering formulation that neglects light reflection and refraction effects.
In our work, we additionally reconstruct a volumetric \ac{IoR} field that along with simulating the laws of physics associated with refractive effects enables us to explain the input RGB images during learning, and consequently provides a meaningful synthesis of novel views at the test time.

\mysubsection{Transparent surface reconstruction}{RWTransparentSurface}
For a survey on reconstruction of shape, illumination and materials of transparent objects, see 
\citet{ihrke2010transparent}.

\citet{kutulakos2008theory} investigate two-interface refractive light interaction with a surface, and for every pixel recover multiple 3D points, so that a ray exiting the surface can be reconstructed.

Environment matting techniques solve for a background deformation by a transparent object, so that it can be composited onto different backgrounds
\cite{zongker1999environment,chuang2000environment,peers2003wavelet,matusik2002acquisition,wexler2002image}.
\citet{khan2006image} and others \cite{yeung2011matting,chen2019learning} demonstrate that even significant departs from physics can be tolerated by human perception to make such compositing look realistic. 

Dedicated setups for transparent object reconstruction rely on light-field background displays \cite{wetzstein2011refractive}, X-ray \ac{CT} scanners \cite{stets2017scene}, and transmission imaging \cite{kim2017acquiring}.
In intrusive setups, which require immersing transparent objects into a liquid with matching \ac{IoR}, straight light paths can be assumed greatly simplifying \ac{CT} reconstruction \cite{trifonov2006tomographic} or range scanning when fluorescent liquid is employed  \cite{hullin2008fluorescent}.

Inspired by environment matting, \citet{wu2018full} and Lyu et al.~\shortcite{lyu2020differentiable} place a transparent object on a turntable in front of a coded background and capture its multiple views from a static camera position. 
\citet{wu2018full} derive the correspondence between the incident (camera) and exit rays that reach the background, which additionally requires rotating the background, and finally consolidate the resulting point clouds into a clean geometric model.
\citet{lyu2020differentiable} perform coarse-to-fine mesh optimization, driven by differentiable tracing of refractive two-bounce light paths, so that distorted refractive patterns and object silhouettes match captured photographs.
Differentiable rendering 
is also employed to \revised{optimize an \ac{IoR} field in order to} cast desired caustics \cite{nimier-david2019mitsuba} or \revised{to design} advanced optical systems that account for optical aberrations \revised{\cite{sun2021end-to-end,tseng2021deep}}. 

\citet{li2020through} employ a cell phone to capture a small number of views that along with segmented transparent object masks and a known environment map are provided as the input for their method.
\citet{sajjan2020clear} show that by employing an RGB-D camera the segmentation task is further simplified.
Similar goals can be achieved using even a single RGB image and a massively trained encoder-decoder network \cite{stets2019single}. 
As pointed out in \citet{lyu2020differentiable} the domain gap can still be expected, as these networks \cite{stets2019single,li2020through,sajjan2020clear} are trained on renderings.

\mysubsection{Eikonal rendering}{RWEikonalRendering}
Light propagation in media with varying \ac{IoR} has been modeled based on formulations derived from the eikonal and transport equations.
Mirage rendering~\cite{berger1990rendering,berger1990ray,musgrave1990note} is concerned with tracing of rays through discrete atmosphere layers, so that the \ac{IoR} increases with elevation.
\citet{stam1996ray} extend this discrete formulation to media with continuously varying \ac{IoR} by introducing the eikonal equation to rendering applications.
\citet{gutierrez2005non} revisited mirages and other atmospheric effects rendering using such continuous formulation.
\citet{ihrke2007eikonal} derived a wavefront tracing technique from the eikonal equation to pre-compute the irradiance distribution in a volume that enables efficient rendering of media with non-homogeneous \ac{IoR}.
Our problem is rather inverse rendering, which has been applied to eikonal forward models in earth sciences \cite{smith2021eikonet} and interferometric tomography  \cite{sweeney1973analytical,liu1989reconstruction,tian2011tomographic} 
to infer 2D or even 3D structure of physical parameters such as velocities, densities or temperatures.

\myfigure{Eikonal}{Emission-absorption (left) and eikonal light transport (right).
Light is the yellow arrow, its thickness indicates strength.
We show three discrete steps. In emission-absorption, direction remains unaltered. In the eikonal formuation, direction changes according to the gradient of the \ac{IoR}, $\nabla\ior$.
In the eikonal case, strength remains unaffected.}

\mycfigure{Overview}{
Overview of the pipeline enabling final eikonal training:
We start by estimating camera poses \cite{schoenberger2016sfm}.
We ask \ac{NeRF} to explain the scene using emission-absorption and straight rays.
In a semi-automated process, we identify a 3D box region not explained and consider this the refractive volume which we exclude from a second \ac{NeRF} fit.
We then grid the view-independent part of this fit to enable the final progressive training using eikonal equations and curved rays.}

\mysection{Light transport ODE Zoo}{Zoo}
We will here discuss three approaches to model interaction of light and matter as \acp{ODE}: 
a complete model (\refSec{CompleteModel}),
an emission-absorption-only model (\refSec{EAModel}) 
and an eikonal-only model (\refSec{EikonalModel}).
The complete one handles refractive and non-refractive scenes, but was only applied to synthetic scenes in the literature.
The emission-absorption one can be used for inverse rendering, but excludes refraction.
Our eikonal one, in combination with the emission-absorption one, can handle refractive transparency in practical inverse rendering.

\mysubsection{Complete model}{CompleteModel}
When light travels through a scene, it changes its radiance \radiance due to absorption and emission as described by the \revised{(refractive)} \ac{RTE} \revised{\cite{preisendorfer1957mathematical}} 
\begin{equation}
\label{eq:RadianceDirectODE}
\revised{n(s)^2}\dds{(\radiance\revised{/n^2})}
=
-\absorption(s)\radiance(s)+\emission(s)
\revised{,}
\end{equation}
\revised{where $n$ is the \ac{IoR} and $s \in [0,\infty[$ is the distance along a (curved) light path, \absorption is the extinction coefficient, and \emission/\absorption is the source function (which includes in-scattering) \cite{chandrasekhar1950radiative}.
The quantity $L/n^2$ is sometimes referred to as basic radiance.
For a spatially varying \ior,} light also changes its position \location and direction \motion due to refraction according to the laws of eikonal light transport \cite{stam1996ray,gutierrez2005non,ihrke2007eikonal}, see \refFig{Eikonal}. \revised{We can describe this using Hamilton's equations for ray tracing~\cite{ihrke2007eikonal}:}
\begin{equation}
\label{eq:PositionDirectODE}
\dds{\location}
=
\frac{\motion(s)}{n(s)}
\text{\qquad and \qquad}
\dds{\motion}
=
\nabla\ior(s)
,  
\end{equation}
\revised{where $\motion$ is not unit length but normalized by $n$.} This model has been used in a virtual setting to render advanced visual phenomena including refraction, total internal reflection, \revised{and scattering~\citep{gutierrez2005non,ihrke2007eikonal,ament2014refractive,pediredla2020path}.}
Unfortunately, this is an ideal model that has not been demonstrated to be tractably used for \ac{NVS} directly.
We will next show the typical simplifications made when ignoring refraction, and introduce a different, also simplified model, that will allow our \ac{NVS} for refraction.

\mysubsection{Emission-absorption-only model}{EAModel}
In \ac{NeRF} \cite{mildenhall2020nerf}, radiance remains subject to emission and absorption
\begin{equation}
\label{eq:RadianceEAODE}
\dds{\radiance}
=
-\absorption(s)\radiance(s)+\emission(s)
,
\end{equation}
but travels along a constant direction \motion and the change of direction is assumed zero (\refFig{Eikonal}-left):
\begin{equation}
\label{eq:PositionEAODE}
\dds{\location}
=
\motion
\text{\qquad and \qquad}  
\dds{\motion}
=
0.
\end{equation}
This is classic ray marching along straight rays \cite{max1995optical}.

\mysubsection{Eikonal-only model}{EikonalModel}
Complementary and finally, we consider a simplified light transport that does not emit or absorb,
\begin{equation}
\label{eq:RadianceEikonalODE}
\dds{\radiance}
=
0,
\end{equation}
but changes direction as per \revised{eikonal light transport} (\refFig{Eikonal}-right):
\begin{equation}
\label{eq:PositionEikonalODE}
\dds{\location}
=
\frac{\motion(s)}n
\text{\qquad and\qquad}
\dds{\motion}
=
\nabla\ior(s).
\end{equation}

\mysubsection{Solving}{Solving}
\mymath{\solverState}{\mathbf z}
Concisely, all three variants can be formulated as position-motion-radiance state \revised{vector} and its derivative:
\begin{equation}
\solverState\revised{(s)}=
\revised{(}\location,\motion,\radiance\revised{)}
\qquad
\text{ and }
\qquad
\solverState'\revised{(s)}=
\dds{\solverState}
.
\end{equation}

For all three approaches, coupled \acp{ODE} 
\begin{equation}
\solverState(s_1)=
\solverState(s_0)
+
\int
_{s_0}
^{s_1}\!\!\!
\solverState'(s)\,
\mathrm d s
=
\mathtt{odeSolve}
(s_0, s_1, \solverState, \solverState')
\end{equation}
need to be solved to compute the final state given the initial state as well as the \ac{IoR}, emission and absorption fields.

Typically, numerical integration such as Euler solvers are used to solve for the state \cite{wanner1996solving}.
Working backwards, to compute gradients of the emission or absorption is done by automatic differentiation of forward Euler solvers  \cite{mildenhall2020nerf,henzler2019escaping}.
Unfortunately, this requires memory in the order of the number of steps a solver takes.
When also accounting for \ac{IoR} with many small steps, this can quickly become prohibitive.
Instead, we use the adjoint \cite{pontryagin1987mathematical} formulation from Neural ODE \cite{chen2018neural,stam2020computing} that uses constant memory also in backward mode to perform $\mathtt{odeSolve}$.

\mysection{Our Approach}{OurApproach}

Our approach has two main steps: First (\refSec{NonEikonalStep}), reconstructing the opaque scene using a non-eikonal emission-absorption model with straight rays (\refSec{EAModel}) and, second (\refSec{EilkonalStep}), modeling the remaining refractive part using an eikonal formulation (\refSec{EikonalModel}).

The result of the first step is an input to the second step, \ie we first train a non-refractive 3D explanation of the world which is input to a second training that 3D-bends rays inside a fixed non-refractive world so that 2D input images can be explained (\refFig{Overview}).

\newcommand{\Rho}{P}
\mymath{\emissionModel}{\bar\emission}
\mymath{\absorptionModel}{\bar\absorption}
\mymath{\iorModel}{\ior}
\mymath{\maskedEmissionModel}{\emission}
\mymath{\maskedAbsorptionModel}{\absorption}
\mymath{\refractiveBox}{\Pi} 
\mymath{\confidence}{c}
\mymath{\errorWeight}{\alpha}
\mymath{\render}{\mathcal R}
\mymath{\emissionGrid}{Q}
\mymath{\absorptionGrid}{\Rho}
\mymath{\gridKernel}{\kappa}
\mymath{\emissionModelParameters}{\theta}
\mymath{\absorptionModelParameters}{\phi}
\mymath{\iorModelParameters}{\psi}

\mysubsection{Non-eikonal step}{NonEikonalStep}
In this step, we train a \ac{NeRF} model of emission (\emissionModel) and absorption (\absorptionModel), assuming straight rays. 
This is used to represent the background and to find the 3D region not explained by the model.
We also learn a multi-scale version of this model to be used in the next step.

\myparagraph{Registration}
In a first step, we compute matrices to transform the camera space of each input image into one reference view using COLMAP \cite{schoenberger2016sfm}.
Hence, we also know the 3D ray for every 2D pixel.

\myparagraph{Diffuse-opaque init}
\hspace{-2pt}Given this information an off-the-shelf \ac{NeRF} is learned that describes emission and absorption as two \ac{MLP}s that fit continuous functions $\emissionModel(\location,\direction)\in\mathbb R^3\times\Omega\mapsto\mathbb R^3$ and $ \revised{\absorptionModel(\location)}\in\mathbb R^3\times\Omega\mapsto\mathbb R$ mapping position and direction to RGB color or scalar opacity.
Let \emissionModelParameters and \absorptionModelParameters denote the \ac{MLP}  parameters of the emission and absorption models resulting from this optimization.

\myparagraph{Masking}
The model above of \emissionModel and \absorptionModel will not be reliable for refractive objects.
Hence, we would like to eliminate these parts of 3D space, and explain them by our eikonal approach.
The parts that are non-refractive, will be input to this step.
We assume the refractive part of the scene can be bounded by a 3D box $\refractiveBox\in\mathbb R^{3\times 2}$ that exclusively contains refractive objects.
This results in a \emph{masked} emission model \maskedEmissionModel, respectively a masked \maskedAbsorptionModel:
\begin{equation}
\revised{\textrm{$\maskedEmissionModel(\location, \direction)$ resp. $\maskedAbsorptionModel(\location)$}}
= \left\{
\begin{array}{ll}
0&
\textrm{\revised{for} $\location \in \refractiveBox$} \\
\revised{\textrm{$\emissionModel(\location, \direction)$ resp. $\absorptionModel(\location)$}}
& \textrm{otherwise}. \\
\end{array}
\right.
\end{equation}

We find the box \refractiveBox \revised{by providing a user with 10 percent of the training images uniformly distributed around the refractive object.} The user selects a few points on the horizontal and vertical extent of the refractive object in the image. Once we have collected these 2D points from the images, we use the \revised{depth map computed from the}  \protect\ac{NeRF} model to find their corresponding 3D locations. We then take 0.02 and 0.98 percentiles of all points along each spatial dimension and multiply them by a constant value of 1.2 to make sure the box encompasses the entire object. The parameters of \refractiveBox are given by the minimum and maximum coordinate values of the points.

\myparagraph{Progressive grids}
Our experiments have shown that solving for the eikonal directly given \maskedAbsorptionModel and \maskedEmissionModel is challenging. The problem is that when rays bend a lot it becomes harder to find correspondences between input images and background. Moreover, the bending depends on the spatial gradient of the \ac{IoR} rather than the \ac{IoR} directly, which is an operation known to be numerically demanding to optimize over.
Addressing this challenge, we will instead learn eikonal transport using different progressively finer versions of the emission and absorption models.
This is inspired by progressive spatial encodings \cite{park2021nerfies}, but instead of blurring the periodic spatial functions, we blur the radiance function itself.

It is not obvious how to make a coarser version of \maskedEmissionModel or \maskedAbsorptionModel which are \acp{MLP}.
In particular, our preliminary experiments using slower-varying or fewer spatial encodings did not result in the desired band-limiting.
Instead, we recur to relying on regular grids.
These are typically struggling to resolve fine details, or to work in 5D, but fortunately, this is not required in our case.
Hence, we sample the masked emission and absorption solutions to a 3D grid as \emissionGrid and \absorptionGrid, collapsing \revised{$\emissionGrid$} over the angular domain:
\mymath{\otherLocation}{\mathbf y}
\begin{eqnarray}
\revised{\emissionGrid_i(\location)}
&=&
\mathbb E_\otherLocation[
\mathbb E_\direction[
\revised{\maskedEmissionModel(\otherLocation, \direction)}
\kappa_i(|\location-\otherLocation|)
]]\\
\revised{\absorptionGrid_i(\location)}
&=&
\mathbb E_\otherLocation[
\revised{\maskedAbsorptionModel(\otherLocation)}
\kappa_i(|\location-\otherLocation|)
]
,
\end{eqnarray}
where $\gridKernel_i$ is a Gaussian kernel of increasing frequency bandwidth for increasing levels $i$.
In our experiments, we use a grid size of $128^3$ and \revised{the values inside the grid are interpolated with a trilinear interpolation scheme. }

\mysubsection{Eikonal step}{EilkonalStep}

At this step, 
we have access to a hierarchy of grids describing the emission and absorption in the scene for all locations $\location\not\in\refractiveBox$ outside the refractive box.
We now find an \ac{IoR} field defined on $\location\in\refractiveBox$ to explain both the non-refractive 3D grids and the 2D images.

\myparagraph{Masked traversal}
\refFig{EnterExit} shows a red ray starting from point $A$ and traversing the world outside \refractiveBox, which is hit at point $B$.
We use the emission and absorption models \emissionModel and \absorptionModel to trace the straight ray from $A$ to $B$ (\refSec{EAModel}).
Starting at $B$, eikonal ray-marching curves out the yellow path (\refSec{EikonalModel}) according to an \ac{IoR} model that maps spatial position to \ac{IoR}:  $\ior(\location)\in\mathbb R^3\mapsto\mathbb R$.
When this ray leaves \refractiveBox at $C$, we let it continue with emission and absorption on a straight path, eventually receiving a contribution at $D$ or other points.

\mywfigure{EnterExit}{0.4}{Enter and exit.}
A key concept is to \emph{enter and exit} the refractive box in a masked traversal, as well as training with masked rays and progressively.

Training of \iorModel\ -- that is also an \ac{MLP} whose parameters are denoted as  \iorModelParameters\ -- proceeds similar to \ac{NeRF}, but instead of marching geometrically, we solve (and back-propagate through) an \ac{ODE} in position-motion-radiance space.

\mymath{\parameters}{\theta}
Recall that we let \solverState denote a position-motion-radiance \revised{vector}. We use dot notation to pick an element in the \revised{vector} so that \solverState.\location denotes the position and \solverState.\radiance the radiance, for example. 
In the mixed refractive/non-refractive case, \revised{the} state \ac{ODE} is 
\begin{equation}
\label{eq:MixedODE}
\solverState_\iorModelParameters'(s)=
 \left\{
\begin{array}{ll}
\text{
\refEqs{RadianceEikonalODE}{PositionEikonalODE}
s.t. }\iorModel_\iorModelParameters&
\text{if }
\solverState_\iorModelParameters(s).\location\in\refractiveBox
\\
\text{
\refEqs{RadianceEAODE}{PositionEAODE}
s.t. }\maskedEmissionModel_\emissionModelParameters \text{ and } \maskedAbsorptionModel_\absorptionModelParameters&
\textrm{otherwise}, \\
\end{array}
\right.
\end{equation}
so the state change is non-eikonal outside the box and eikonal inside.
It is made to depend on \iorModelParameters, but not on \emissionModelParameters and \absorptionModelParameters, as these are fixed both in the forward and backward pass of this step.

Let \revised{$\solverState_i$} denote the state of a ray \revised{through} pixel $i$.
We then find
\begin{equation}
\iorModelParameters^\star=\operatorname{arg\,min}_{\revised{\iorModelParameters}}
\mathbb E_i
[
|
\mathtt{odeSolve}
(s_0,s_1, 
\solverState, 
\solverState',
\iorModelParameters).\radiance
- \solverState_i.\radiance
|
]
,
\end{equation}
where \iorModelParameters is an extra argument for $\mathtt{odeSolve}$ with parameters that condition $\solverState'$.

As a ray cannot change direction outside \refractiveBox, the condition in \refEq{MixedODE} can be handled by loop splitting in practice: First the ray is traced straight, then traced eikonal, and then it is traced straight once more, eliminating the conditional statement in \refEq{MixedODE}.

\myparagraph{Masked rays}
Since our \ac{MLP} for estimating the \ac{IoR} is only evaluated inside the bounding box, we start the eikonal training by making sure a batch contains only the rays that are hitting the box.

\myparagraph{Progression}
We start by finding an \ac{IoR} field that explains a coarse version of the emission-absorption grid.
When the change of error falls below a threshold, we switch one level up to a finer grid.
The number of parameters in the \ac{MLP} to represent the \ac{IoR} is the same at all levels.
We render final images using the full \ac{NeRF} model instead of a grid.

\myparagraph{Interior radiance field}
Non-transparent objects might be present in the interior of the transparent object that we located in \refractiveBox. To explain these, we train another \ac{NeRF} for the radiance in \refractiveBox. The \ac{IoR} field in \refractiveBox is available from the eikonal step (\refSec{EilkonalStep}), and we can now keep it fixed together with the opaque \ac{NeRF} (\refSec{NonEikonalStep}) and trace paths that bend according to the eikonal when encountering the transparent object in \refractiveBox. Conclusively, our solution consists of the opaque \ac{NeRF}, the \ac{MLP} for the \ac{IoR} field, and a \ac{NeRF} for the interior of the transparent object. Together, these have been trained sequentially to explain the input images.

\mysubsection{Implementation details}{Details} Our \ac{NeRF} implementation follows \citet{mildenhall2020nerf}, and our second \ac{MLP} to represent the \ac{IoR} field is a 6-layer \ac{MLP} with 64 hidden dimensions with a skip connection that concatenates the input to the third layer’s activation.
Similar to \ac{NeRF}, we also apply positional encoding with five frequencies to the input.
For stable training, as suggested by \citet{chen2018neural}, we use softplus activation with $\beta=5$ for all layers instead of a non-smooth function like ReLU \revised{and all layers are initialized with the Xavier uniform.} 
In the non-eikonal step (training \ac{NeRF}), we use the same training setting as described by \citet{mildenhall2020nerf}, and let the optimization run for 150k iterations. This takes around 12 hours to converge on a single NVIDIA 1080Ti with 12 GB RAM. For the eikonal step, we use a batch size of 1024 rays and traverse the space with 128 ODE steps and the training takes around 5 hours for 5k iterations. \revised{We use the Neural ODE PyTorch implementation \cite{chen2018neural,stam2020computing} to backpropagate through the ODE with the adjoint method.} In the progression part, we smooth the grid with a Gaussian kernel with a \revised{normalized frequency bandwidth of 0.08 cycles per sample} and double this for every 1k iterations.
For the last step, we use a single MLP similar to the \ac{NeRF} fine network \cite{mildenhall2020nerf}, but with 128 hidden dimensions to represent the interior radiance field.
As we do not adopt any hierarchical volume sampling, we consider 512 steps along the ray to properly sample both interior and exterior radiance fields, and it takes around 12 hours to optimize over 10k iterations. 
With our complete model, it takes around 85 seconds to render a frame of 672$\times$504 resolution. 

\newcommand{\method}[1]{\textcolor{color#1}{\texttt{\textbf{#1}}}}

\colorlet{colorNeRF}{colorA}
\colorlet{colorTrivial}{colorB}
\colorlet{colorTriv\-i\-al}{colorB}
\colorlet{colorOurs}{colorC}
\colorlet{colorDirect}{colorD}

\newcommand{\scene}[1]{\textsc{#1}}

\newcommand{\winner}[1]{\fontseries{b}\selectfont{#1}}
\begin{table*}[htb]
    \setlength{\tabcolsep}{3.6pt} 
    \centering
    \caption{Quantitative comparison of different methods (rows) using different scenes and metrics (columns).
    The numbers in the User column say how often in our user study the method was considered closer to the reference than ours. As these numbers are significantly ($p < 0.01$) smaller than the chance level 50\%, our method was for all scenes considered closest to the reference in the majority of the comparisons shown to the users.
    }
    \sisetup{
      detect-weight=true,
      detect-inline-weight=math,
      table-number-alignment = center,
      table-align-text-post = false,
      output-decimal-marker=\textmd{.},
    }
    \vspace{-2ex}
    \label{tab:Results}
    \begin{tabular}{l 
        S[table-format=2.3]
        S[table-format=1.3] 
        S[table-format=1.3] 
        S[table-format=1.3]
        S[table-format=2.3]
        S[table-format=1.3]
        S[table-format=1.3] 
        S[table-format=1.3]
        S[table-format=2.3]
        S[table-format=1.3]
        S[table-format=1.3]
        S[table-format=2.3]
        S[table-format=2.3]
        S[table-format=1.3]
        S[table-format=1.3]
        S[table-format=2.3]
        }
        \toprule
        &
        \multicolumn4c{\scene{Ball}}&
        \multicolumn4c{\scene{Glass}}&
        \multicolumn4c{\scene{Pen}}&
        \multicolumn4c{\scene{WineGlass}}
        \\
        \cmidrule(lr){2-5}
        \cmidrule(lr){6-9}
        \cmidrule(lr){10-13}
        \cmidrule(lr){14-17}
        &
        \multicolumn1c{PSNR}&
        \multicolumn1c{SSIM}&
        \multicolumn1c{LPIPS}&
        \multicolumn1c{User}& 
        \multicolumn1c{PSNR}&
        \multicolumn1c{SSIM}&
        \multicolumn1c{LPIPS}&
        \multicolumn1c{User}& 
        \multicolumn1c{PSNR}&
        \multicolumn1c{SSIM}&
        \multicolumn1c{LPIPS}&
        \multicolumn1c{User}& 
        \multicolumn1c{PSNR}&
        \multicolumn1c{SSIM}&
        \multicolumn1c{LPIPS}&
        \multicolumn1c{User} 
        \\
        \midrule        
\method{NeRF} & \bfseries 27.384 & 0.945 &  0.042 & 0.27 \% &  \bfseries27.146 & \bfseries 0.924 & 0.066 & 3.83 \% &  27.749 & 0.933 & 0.059 & 9.58 \% &  \bfseries 29.011 &  \bfseries 0.947 & 0.045 & 24.93 \% \\
\method{Trivial} & 24.373  & 0.933 & 0.034 & 9.31 \% & 25.930 & 0.914 & 0.059 & 7.39 \% & 23.070 & 0.912 & 0.060 & 37.26\si{\percent} & 26.739 & 0.935 & 0.052 & 1.33 \% \\
\method{Direct} & 27.247 & 0.942& 0.054 &  & 26.031 & 0.914& 0.081 & & 26.624&0.928 & 0.073&  & 27.379 & 0.938& 0.060 & \\

\method{Ours} & 26.720 & \bfseries 0.951 &  \bfseries 0.023 & &  26.525 & 0.922 & \bfseries 0.050 & & \bfseries 27.803 & \bfseries 0.935 & \bfseries 0.047 & & 27.789 & 0.940 & \bfseries 0.042 &   \\
        \bottomrule
    \end{tabular}
\end{table*}

\mysection{Results}{Results}
Our aim is \ac{NVS} with plausible coherence in scenes with transparent objects.
We look at a range of scenes where we apply different methods and -- as no metric exists to quantify our main aim -- we refer to standard \ac{PSNR}, \ac{SSIM}, and \ac{LPIPS} metrics, and we perform a user study too.

\myparagraph{Scenes}
We selected four real scenes including refractive objects with unknown geometry: 
\scene{Ball}, \scene{Glass}, \scene{Pen} and \scene{WineGlass}.
We used an Iphone 8 camera to capture  96, 97, 105, and 102 views, respectively for each scene, and we hold out 1/10 of all views for the test set. All images are of resolution 672$\times$504 pixels.

\myparagraph{Methods}
We compared \method{Ours} with \method{NeRF} and \revised{two other methods named \method{Direct} and \method{Trivial}}.
\revised{In \method{Direct}, we jointly optimize for the emission-absorbtion and \ac{IoR} models. In this setup, similar to progressive grid, we provide a coarse-to-fine optimization scheme by applying progressive positional encoding \cite{park2021nerfies} for the emission-absorbtion \ac{MLP}.} In \method{Trivial}, we try to reconstruct the \ac{IoR} field using the density field of refractive objects recovered by \ac{NeRF}.
We do this by first executing the \ac{NeRF} model for a discrete set of samples along the rays \revised{coming from the input camera poses and} crossing the bounding box \refractiveBox, and we set the density to zero for the samples outside the box. Then, for each ray, we \revised{estimate both the front and back surface position of the refractive object by forward and backward ray marching until an opacity threshold is reached (similar to how the depth maps are computed in \ac{NeRF}).}
For the samples that fall between the intersections, we assign a constant \ac{IoR} value (1.5 for \scene{Glass}, 1.33 \scene{WineGlass}), and we choose 1.0 for the regions outside. We then try to fit an \ac{MLP} to map each 3D point inside the box \refractiveBox to its calculated \ac{IoR}.

\myparagraph{Qualitative comparisons}
\refFig{Results} facilitates a visual comparison of \method{Ours} with \method{NeRF} and \method{Trivial}. \revised{Please refer to the supplemental material for our visual comparison with the \method{Direct} method.}
The insets show novel view reconstructions of different view points for all methods.
Please refer to the supplemental video for an animated version of these results.
\method{NeRF} tends to ``fake'' refraction by considering a diffuse content on the surface of the transparent object and assigning view-dependent color for each point on the surface. 
Under the condition of extreme view changes, as can be seen in all scenes, \method{NeRF} fails to properly reproduce the color and it tends to average all observations leading to a blurry result. 
\method{NeRF} also seems to struggle with reconstruction of an occluder inside the transparent objects although multi-view consistency holds for the object inside. In the \scene{Pen} scene, \method{NeRF} failed to assign a transparent content on the surface of the glass in order to properly reconstruct the pen inside.
\method{Trivial} assumes a constant \ac{IoR} field inside the entire refractive object and in case of spatially varying \ac{IoR}, the refraction tends to be wrong for some regions (\eg towards the top and the bottom of the glass in the \scene{Glass} and the \scene{Pen} scenes). \method{Trivial} performs better on the \scene{Ball} scene as the crystal ball has a constant \ac{IoR} inside. However, due to the mere fact that the \ac{NeRF} density field for the refractive object is not always valid, the estimated \ac{IoR} of \method{Trivial} might not be very accurate and the refracted background becomes misplaced in some regions. 
In contrast, \method{Ours} reproduces sharper details and aligns better with the reference. 
Moreover, in order to assess the temporal consistency of each method, in the right block, we also show the corresponding pseudo-epipolar image that is created by stacking a selected scanline for 30 subsequent video frames using a continuous camera trajectory.
A good optical flow continuity can be observed between the stacked scanlines for all methods, but clearly the flow fidelity with respect to the reference is best for \method{Ours}. 
\method{NeRF} and \method{Trivial} feature significant blur that is visible also in the insets in the middle column.

\mycfigure{Results}{\revised{The left block shows the cross section of recovered \ac{IoR} by our method for a scanline between the white dots shown in the reconstructed test view in the second block.}
The third block shows insets taken from novel views produced by three different methods (rows) for different view points (columns).
The right block shows a pseudo-epipolar view using a continuous camera trajectory, again for all methods.}

\myfigure{Synthetic}{\revised{\ac{IoR} cross section of our method (orange) and ground truth (blue) for a scanline along the pixel marked with dot in each inset.}}

\myparagraph{User study}
Unfortunately, no method exists to quantify the main aim of this work, plausible refractive and reflective flow.
To quantify the coherency, we performed a small user study.
\revised{A reference photograph and two images produced by \method{Ours} and either \method{NeRF} or \method{Trivial} (selected randomly) were shown to} 73 participants, \revised{10 image triplets for each scene}.
The participants then had to indicate which one is visually closer to the reference in a \ac{2AFC} experiment.
All three images were presented simultaneously without any time limit; the position of the reference was fixed, while it was randomized for the other two.
We selected five different views for each of the four scenes 
and aggregated the participant selection over those views.
For each scene, we report how often a competitor was selected, hence less is better \revised{for us}, while the chance level is 50\%.
All outcomes are significant at the $p<0.01$-level for a binomial test at $N=73$.

\myparagraph{Quantitative comparisons}
\refTab{Results} presents quantitative results of our user study  \revised{(in the ``User'' columns)} and for the different metrics averaged over our test set.
We see \method{NeRF} has consistently the highest \ac{PSNR}, which is a metric relatively insensitive to blur or structure preservation.
When it comes to \ac{SSIM}, already a metric more aware of the structures we want to preserve, it comes to a draw.
At the most advanced metric, \ac{LPIPS} -- which is based on human image artifact perception and better tolerates small spatial misalignments with respect to the reference -- \method{Ours} always wins.
\revised{\method{Direct} and \method{Trivial} are sometimes better than other methods but never win.}
The participants of our user study almost consistently indicate that \method{Ours} leads to less perceived differences with respect to the reference views.
As for a relatively high score of \method{Trivial} for \scene{Pen}, we hypothesize that the background sharpness and its color saturation could have appeal to some  participants, who neglected strong background distortions, and the pen's absence, visible in \refFig{Results}.
The relatively high score of \method{NeRF} for \scene{WineGlass} can be attributed to views where the background contained less high-frequency details, so that blur became perceivable.

\myparagraph{Reference comparison.}
\revised{While our method makes use of an \ac{IoR}, it is not forced to use actual physical values.
For a scene with a known \ac{IoR} (\refFig{Synthetic}), we see that it can reconstruct the image faithfully, while a cross-section shows the \ac{IoR} is indeed quite different from the reference \ac{IoR}. We would hence like to iterate that our method is suitable for \ac{NVS}, not for the reconstruction of 3D structure.}

\mysection{Conclusion}{Conclusion}
Given a set of 2D images containing refractive materials, we explored the problem of optimizing for the field of 3D-spatially varying \ac{IoR} with the purpose of \ac{NVS}.
Existing solutions that learn 3D fields for \ac{NVS} are based on opaque or transparent light transport along straight paths.
As opposed to this, we model the bending of light according to the eikonal equation from geometric optics.
This enables us to do perceptually better \ac{NVS} in 3D scenes with complex objects exhibiting strong refractive effects.

Our work is subject to several assumptions.
The eikonal equation deals with refraction and total internal reflection but not separation into partial reflection and refraction.
Partial reflection and refraction in continuously varying media is difficult even in forward simulation, and left for future work.
We also first learn the diffuse world, followed by the transparent objects in a second pass, where we rely on a user marking the bounding box of the specular object to aid the task.
Ideally, this would be done jointly and in a fully automated way.
As a consequence, we have to assume that we sufficiently observe the diffuse world directly, making us unable to reconstruct parts exclusively revealed in the refraction.

Despite our simplifying assumptions, we have by means of eikonal light transport for the first time included refraction and total internal reflection in a model that learns 3D fields from images of transparent objects to accomplish synthesis of novel views. We compared our results with other methods and conducted a user study which strongly indicates that we achieve results that are perceptually closer to reference images.


\bibliographystyle{ACM-Reference-Format}

\end{document}